\documentclass[preprint,showpacs,preprintnumbers,amsmath,amssymb,%
               superscriptaddress]{revtex4}

\usepackage{graphicx}
\usepackage{dcolumn}
\usepackage{bm}

\begin{document}

\preprint{\today, Version 01}

\newcommand{\bb}{{\bf b}}
\newcommand{\beq}{\begin{equation}}
\newcommand{\eeq}{\end{equation}}
\newcommand{\bB}{{\mathbf{B}}}
\newcommand{\bE}{{\bf E}}
\newcommand{\bH}{{\bf H}}
\newcommand{\bD}{{\bf D}}
\newcommand{\bM}{{\bf M}}
\newcommand{\bN}{{\bf N}}
\newcommand{\bL}{{\bf L}}
\newcommand{\br}{{\bf r}}
\newcommand{\bV}{{\bf V}}
\newcommand{\murb}{{\bar{\mu}_r}}
\newcommand{\etheta}{{\bf e}_{\theta}}
\newcommand{\ephi}{{\bf e}_{\phi}}
\newcommand{\er}{{\bf e}_{r}}
\newcommand{\ex}{{\bf e}_{x}}
\newcommand{\ey}{{\bf e}_{y}}
\newcommand{\ez}{{\bf e}_{z}}
\newcommand{\ep}{{\bf e}_{+}}
\newcommand{\emm}{{\bf e}_{-}}
\newcommand{\eo}{{\bf e}_{0}}
\newcommand{\uvmn}{\,^{uv}_{mn}}
\newcommand{\mnuv}{\,^{mn}_{uv}}
\newcommand{\mnpq}{\,^{mn}_{pq}}
\newcommand{\dsum}{\displaystyle\sum}
\newcommand{\wtc}{\widetilde{c}}
\newcommand{\wtd}{\widetilde{d}}
\newcommand{\barc}{\bar{c}}
\newcommand{\bard}{\bar{d}}
\newcommand{\bark}{\bar{k}}
\newcommand{\tg}{\widetilde{g}}
\newcommand{\te}{\widetilde{e}}
\newcommand{\tf}{\widetilde{f}}
\newcommand{\barg}{\bar{g}}
\newcommand{\bare}{\bar{e}}
\newcommand{\barE}{\bar{E}}
\newcommand{\barf}{\bar{f}}
\newcommand{\breg}{\breve{g}}
\newcommand{\bree}{\breve{e}}
\newcommand{\bref}{\breve{f}}
\newcommand{\cG}{{\cal G}}
\newcommand{\cE}{{\cal E}}
\newcommand{\cF}{{\cal F}}

\title{Critical velocities and the effect of steady and oscillating
rotations on solid He4}

\author{S T Chui}
\affiliation{Bartol Research Institute, University of Delaware,
        Newark, Delaware 19716}

\date{\today}

\begin{abstract}
We apply our recently developed model of a Bose condensate
of quantum kink wave in solid He4
to understand recent torsional oscillator
experimental results of the citical velocities and the effect
of the steady and oscillating rotations at around 0.1 degree K. 
When the D.C. rotation is present we find a decrease
of the Q factor given by $Q^{-1}
\propto f_{sf}\times \Omega_{D}/\omega_{TO}$ where
$f_{sf}$ is the superfluid fraction; $\Omega_{D}$, the
D. C. angular rotation velocity, $\omega_{TO}$, the torsional
oscillator oscillating frequency.
We estimate the AC critical velocity $\Omega_A^{crit}$
as that required to generate a kink wave of wavevector
$2\pi/L_d$ where $L_d$ is the distance between nodes of the
dislocation network.
We generalize this to include a steady rotation and 
find a D. C. critical velocity 
$\Omega_D^{crit}
\propto (\Omega_{A}^{crit})^{1/2}$. Estimates for both the
steady and the oscillating critical velocities are in order
of magnitude agreement with experimental results. We have also 
examined an alternative mechanism of kink tunnelling through
a node in the dislocation networm and  find that
there is also a dependence
on the torsional oscillator frequency:
$\Omega_D^{crit}=[\Omega_A^{crit} \omega_{TO}2\pi]^{1/2}.  $
The DC critical velocity $\Omega_D^{crit}$ 
is ten times higher than the experimental value.

\end{abstract}

\pacs{
67.80.-s
}

\maketitle
Since the discovery of an increase (1 per cent)
in the solid $^4$He moment of inertia
in 
torsional oscillator (TO) experiments 
at around 200 mK\cite{chan}, there have been renewed interests in its
low temperature physical properties\cite{x,jb,balibar,reppy,toner,lagb}. Many novel physical behavior are manifested, such as a very small
direct flow and a very small critical velocity ($\Omega_A^{crit}
\approx 10^{-3}$ rad/s).
Recently TO experiments are carried out in the presence of both a 
steady (DC) and an oscillating (AC) rotation\cite{Choi,Kubota}. An 
increase in damping is observed which increases with the DC rotation
speed. When the AC rotation velocity is below the
critical value, there is also a DC critical velocity 
$\Omega_D^{crit}$ which is three orders of magnitude larger than the 
AC critical velocity. The DC rotation does not affect the shear modulus. 
The AC speed of the Kubota group\cite{Kubota}, $60 \mu m$, 
is higher than $\Omega_A^{crit}.$ No critical DC velocity was observed.

We have recently
studied the physics of kink waves of dislocations of density $n_d$ and 
their Bose-Einstein condensation (BEC) \cite{kwave}.
The BEC of the kinks makes possible dissipationless
movement of the dislocation lines.
The motion of a dislocation corresponds in part to a circular 
motion of many He4 atoms, each by a different amount. An estimate
of the fraction of He4 atoms can be obtained by weighting with respect
to the strains. With this the
corresponding "superfluid fraction" due to the motion of the
dislocation lines is found to be of the order of $n_da_0L_m$,
a magnitude that is consistent with current experimental results.
Here $a_0$ is the lattice constant, $L_m$ is the
mosiac size. The dislocation motion does not produce any
net linear motion of the atoms and thus will not generate any direct 
superflow. In this paper we apply our model to understand the
experimental results of the citical velocities and the effect
of the DC and AC rotations.
We estimate the AC critical velocity $\Omega_A^{crit}$
as that required to generate a kink wave of wavevector
$2\pi/L_d$ where $L_d$ is the distance between nodes of the
dislocation network.
When the DC rotation is present we find a decrease
of the Q factor given by $Q^{-1}
\propto f_{sf}\times \Omega_{D}/\omega_{TO}$ where
$f_{sf}$ is the superfluid fraction; $\Omega_{D}$, the
D. C. angular rotation velocity, $\omega_{TO}$, the torsional
oscillator oscillating frequency.
We have examined two 
mechanisms for a DC critical velocity $\Omega_D^{crit}$:
(1) The DC rotation generates kinks with a time
independent displacement. Oscillating kinks are in turn generated 
from this state by the oscillating rotation.
(2) Similar to the Josephson effect,
the combination of the DC and AC rotation can cause a steady 
current of kinks
across nodes of the dislocation network when the DC rotation
is fast enough.  We find that for both mechanisms $\Omega_D^{crit}
\propto (\Omega_{A}^{crit})^{1/2}$ where
$\Omega_{A}^{crit}$ is the critical AC angular velocity. 
For the second mechanism, there is also a dependence
on the torsional oscillator frequency:
$\Omega_D^{crit}=[\Omega_A^{crit} \omega_{TO}2\pi]^{1/2}.  $
Using current experimental estimates
for the different physical parameters, we find $\Omega_A^{crit}$
of the same order of magnitude as the experimental value.   
An estimate of the DC critical velocity $\Omega_D^{crit}$ 
with the first mechanism is 
of the same order of magnitude as the experimental results; with
the second mechanism the critical velocity
is ten times higher than the experimental value.
We hope this paper will stimulate further experiments
and provide tests of the validity of
our picture. We now describe our results in detail.

As is well known\cite{LL}, for a body of density $\rho$
rotating with angular
frequency $\Omega$, the quantity of interest is
\beq
F=E-\Omega\cdot {\bf M}
\eeq 
where $E$, {\bf M} are
the energy and the angular momentum measured with respect
to a coordinate system fixed in space.
For example, for a simple rotation at frequency $\Omega$, consider
a body rotating at velocity $\nu$. Then
$E(\nu)=0.5\int d{\bf r}\rho \nu^2r^2$ 
and $M=\nu\int d{\bf r}\rho r^2$. 
Minimizing $F$ with respect to $\nu$ we get $\nu=\Omega$,
as we expected. At this frequency 
$F(\nu=\Omega)=F_0=-0.5\Omega^2\int d{\bf r}\rho r^2$

In the presence of only a time dependenct oscillating rotation with
an angular frequency ${\bf \Omega}_{TO}(t)=\Omega_A\exp(i\omega_{TO}t)$
caused by a torsional
oscillator, if the dislocations also move with the entire solid, the
energy of the system would be
$E_0=0.5\int d{\bf r}\rho [\Omega_{TO}(t)
\times ({\bf r+ u})]^2.$ Here ${\bf u}$ is the displacement
due to the dislocations.
In our picture, as the He4 is rotated , the kinks remain in
the zero momentum condensate relative to a space fixed coordinate
system. This reduces the kinetic energy of the system. 
We obtain
$E'=0.5\int d{\bf r}\rho [\Omega_{TO}(t)\times
({\bf r+u+ \Delta u})]^2.$ Here ${\bf \Omega}_{TO}(t) \times 
{\bf\Delta u}$ is 
the reduction in the angular velocity  where  ${\bf \Delta u}$
comes from the motion of the dislocations relative to the
rotating  solid. The "superfluid fraction" is given by
$f_{af}=(E_0-E')/E_0.$ 

We first estimate the critical
velocity when only the time dependent oscillation 
with angular velocity ${\bf \Omega}_{TO}(t)$ is present.
We consider that a critical velocity is reached when it becomes
possible to excite a kink wave so that it is possible to lower 
$F$; $\Delta F$ becomes negative. For a network of dislocations
the lowest wavevector is of the order of $k_0=2\pi/L_d$ where $L_d$
is the distance between nodes. This wave vector can be further 
increased if the dislocation moves close to defects (He3) 
which provide further pinning.  In that case, the critical velocity
will become higher. We think a lot of the recently observed
hysteretic behaviour\cite{kojima} is related to this issue.

An example of a kink wave oscillating with frequency $\omega_{TO}$ is 
given by: 
\beq
|\psi>=\sin\omega_{TO} t/2\  |-k_0>+\cos\omega_{TO} t/2\  |k_0>.
\eeq
The average velocity of  this state is given  by
\beq
v=<\psi|{\hat v}|\psi>= v_0\cos\omega_{TO} t
\eeq
where
\beq
v_0=\hbar k_0/m^*,
\eeq
$m^*$ is the effective mass of the kink wave.
We next procced to estimate $\Delta F$.

If there are N kinks per unit length, each with velocity $v$
the velocity at which the dislocation moves out
can be estimated as follows. For the dislocation to move out by a
lattice constant a, each kink has to move a distance of 1/N. The
time it takes to do this is $\Delta t=1/(Nv)$. Denoting the
position of a dislocation by $c$, the speed
of the dislocation due to the finite speed of the kink is 
\beq
\partial c/\partial t 
= a\exp(i\omega_{TO} t)/\Delta t=a\exp(i\omega_{TO} t)Nv.
\label{vc}
\eeq
A dislocation with the Burger's vector along the x direction
at some point $(c_x, c_y)$ along the z axis causes
an atom at position (x, y) to move by $u_x({\bf r-c})$,
$u_y({\bf r-c})$ \cite{uxy}.
The atomic displacement
depends on the position of the dislocation.
\beq
\partial {\bf u}/\partial t = -\nabla_c {\bf u}\cdot
\partial {\bf c}/\partial t.
\label{ut}
\eeq
is the corresponding velocity of an atom a distance r' away from
the moving dislocation due to the finite speed of the kinks. 
In the following, we shall assume that the dislocation
moves in the direction along the Burger's vector which we
take to be the x axis.
In general, the axis of rotation is not parallel to the axis of the
dislocation. With respect to the rotation axis, the actual displacement
should be ${\bf R u(R^{-1}r})$ where {\bf R} is the rotaion
matrix that can be specified by the Euler angles. We shall assume
that that this is the case and for simplicity of notation, not
displayed this dependence at every step.


When the kink wave is created, the kinetc energy cost for a segment
of the dislocation between nodes is given by\cite{ra}
\beq
\Delta E = NL_d(\hbar^2k_0^2/2m^*)
\eeq
where $\delta {\bf u}$ is the displacement
caused by this state. The term $\int d{\bf r}\rho 
(\partial u/\partial t)^2$ has already been included in the kinetic 
energy of the kink\cite{eqn} and thus need not be counted twice.

%

The change in the angular momentum due to a change of the state
($r>>u$)
of the kinks is given by
$$
\Delta {\bf M}
\approx \int d^2 r\rho\ 
{\bf r\times \partial u(r')/\partial t
} 
$$
There is another term $\int d^2r \rho {\bf r\times 
(\Omega_{TO}(t)\times \delta u)} $ which provides a zero time average
to $\Omega_{TO}(t)\Delta M$ and thus will be ignored from now on.
From eqs. (\ref{vc}) and (\ref{ut})
$\Delta {\bf M}$ is of the order
\beq
\Delta {\bf M}
\approx \int d^2 r\rho\ {\bf r}\times amNvL_d
\partial_{x'} {\bf u(r') } 
\eeq
In general {\bf u}  is a sum of contributions from different dislocations
located at different positions ${\bf c}_i$ : ${\bf u}=
\sum_i {\bf u_0(r'-c}_i).$ $\Delta {\bf M}$ can be written as a sum of 
contributions from each of the dislocations. 
\beq
\Delta {\bf M}
\approx \sum_i \int d^2 r\rho {\bf (r-c}_i)\times amNvL_d
\partial_{x'} {\bf u(r'-c_i)} 
\eeq
The range of integration of each of these terms is of the order of
the mosiac size. Since $\partial_{x'} {\bf u(r'-c_i)}$ is
of the order of $1/|r'-c_i|$ 
for a single dislocation, we obatin
\beq
\Delta {\bf M}\approx 2\pi m^*NvL_d L_m^2/a
\eeq
where $L_m$ is the mosiac size.
From eq. (7), (1) and the condition that $\Delta F=0$,
we get the critical angular velocity 
\beq
\Omega_A^{crit}=\Delta E/\Delta M 
\approx \hbar a/(2\pi m^*L_dL_m^2).
\eeq
Using experimental estimates of
$L_d=5\mu m,$
$L_m=20\mu m,$ and our estimate $m^*\approx 0.1 m_{He4}.$
%
%
We get
$\Omega_A^{crit}
\approx 10^{-3}/s,$
of the same order of magnitude as the experimental results. 

%
We next consider the case where the solid is rotating with a 
constant angular velocity $\Omega_D$ and ask if it is energetically 
favorable to start moving the kinks to a state of finite momentum.
Instead of an "oscillating state" as in eq. (2), we consider 
the possibility of creating simple states $|\pm k_0>.$ 
The velocity of the kinks will then just be $v_0$ instead of $v.$
Going through
the same algebra, we arrive at a DC critical velocity that is the
same order of magnitude as the AC critical velocity.
The experimental
DC angular velocity is higher than the
AC angular velocity by two orders of magnitude. 
We thus assume a state so that the dislocations
move with the entire solid with the constant angular velocity
$\Omega_D$. In the additional presence of an oscillating
driving term so that the total angular velocity is
$\Omega=\Omega_D+ \Omega_{TO}(t),$
we now consider if the dislocations will
exhibit oscillating movements. 

If the kinks do not exhibit the oscillating motion,
the kinetc energy saved is
given by 
$$\Delta E=\int d{\bf r}\rho {\bf v}_0\cdot \delta {\bf v}.$$
Here ${\bf v_0=[\Omega_D+\Omega_{TO}(t)]\times r}$ is the velocity
of the solid in constant rotation. 
$\delta {\bf v}(t)$ is the change in velocity due to the 
kinks not moving with an oscillating velocity 
so that the core position $\delta c(t)$ exhibits a oscillating
time dependence relative to the rotating solid.
The change in velocity now has an additional contribution from the 
coupling of the steady rotation:
\beq
\delta {\bf v}=-\nabla_c {\bf u} \cdot (
\Omega_D\times \Delta c+\partial \Delta {\bf c}/\partial t).
\label{vd}
\eeq
$\partial \Delta {\bf c}/\partial t\approx -{\bf \Omega}_{TO}(t)
\times {\bf c},$ 
$\Delta {\bf c}\approx -{\bf \Omega}_{TO}(t)
\times {\bf c}/i\omega_{TO}.$ 
Because ${\bf v}_0$ is a sum of two terms, 
$\Delta E=\Delta E_D+\Delta E_A$ 
contains two contrbutions:
those from copupling to 
$\Omega_D$ and those from coupling to $\Omega_{TO}(t).$
The coupling term to the {\bf constant} DC rotation is given by
$$\Delta E_D=\int d{\bf r}\rho {\bf \Omega_D\times r}
\cdot \delta {\bf v}(t)$$ which has a zero time average.
Thus the DC rotation cannot directly drive the dislocations
to a finite oscillating velocity.
The coupling term to the {\bf oscillating} rotation is given by
$$\Delta E_A=\int d{\bf r}\rho {\bf \Omega_{TO}(t)\times r}
\cdot \delta {\bf v}(t)
,$$
Because $\delta v$ is a sum of two terms (eq. \ref{vd}),
$\Delta E_A=\Delta E_{A1}+\Delta E_{A2} $ where
$\Delta E_{A1} =\int d{\bf r}\rho {\bf \Omega_{TO}(t)\times r}
(\nabla_c {\bf u}\cdot 
\partial \Delta {\bf c}/\partial t),  $ 
$\Delta E_{A2} =\int d{\bf r}\rho {\bf \Omega_{TO}(t)\times r}
(\nabla_c {\bf u} \cdot \Omega_D\times \Delta c).  $ 
$\Delta E_A$ has a nonzero time average.
The ratio $\Delta E_{A1}/E\approx f_{sf}$ provides for the effective 
reduction of the moment of inertia and is of 
the order of the "superfluid fraction" $f_{sf}$. 
Now $\Delta {\bf c}$  and
$\partial \Delta {\bf c}/\partial t$ in eq. (\ref{vd}) and hence
$\Delta E_{A1}$ and $\Delta E_{A2}$ are ninty degree 
out of phase in time.
We thus expect $\Delta E_{A2}$ to provide for a damping term,
as is observed in the experiments. 
The ratio $\Delta c/(\partial \Delta c/\partial t)$ is of the order 
of $1/\omega_{TO}$,
the inverse torsional oscillator vibration frequency.
The total rotation energy $E_T$ of the system is a sum of the
rotation energy of the container and that of solid He4, $E_0$.
The $Q$ factor is defined with respect to $E_T$.
We write $E_0=\alpha E_T$ for a constant $\alpha$.
We thus expect the $\Delta E_{A2}$ term to provide a
damping that is of the order of $E_T/Q$ where
\beq
1/Q\approx \alpha f_{sf}\Omega_D/\omega_{TO}.
\eeq
Taking a superfluid fraction of the order of 1 per cent,
a $\Omega_D$ of the order of 1 rad/s and $\omega_{TO}=2\pi\times 10^3 
rad/s$, we obatin an estimate of $Q$ that is of the order of $10^6
\alpha.$
Experimentally, $Q^{-1}$ ranges from $10^{-6}$ to $10^{-9}.$
Our estimate is consistent with this.
$1/Q$ scales with $\Omega_D$ and $f_{sf}$, 
also consistent with experimental
findings. We next examine the critical DC rotation field.
We have considered two possible mechanisms. We describe them 
sequentially next.

(i) We again examine the energetics of creating a kink wave of 
wavevector $2\pi/L_d$.
Before the kink wave is created, the atoms are at positions
$r_i+u_i+\Delta u_i$. Because $\Delta u_i<<u_i,$ we shall neglect 
the contribution due to $\Delta u_i$ below.
The angular momentum is now given by 
$$
M=m\sum_i [{\bf r_i+u_i+\delta u_{i}}(t)]\times \left (
[{\bf \Omega_D+\Omega_{TO}}(t)]\times [{\bf r_i+u_i+\delta u_{i}}(t)]
+ \partial {\bf u}_{i}/\partial t]
\right ) 
$$
The  velocity ${\bf v}_{i}$
is a sum of that due to motion of the kink, 
$\partial u_{ti}/\partial t,$ and that due to the rotation
$\Omega_D+\Omega_{TO}$  
%
The corresponding energy is
$$
E=0.5m\sum_i\left (
[{\bf \Omega_D+\Omega_{TO}}(t)]\times [{\bf r_i+u_i+\delta u_{i}}(t)]
\right )^2 +NL_d\hbar^2/(2mL_d^2).
$$
Recall that before the kink wave is created, the energy is
$$
E_0=0.5m\sum_i\left (
[{\bf \Omega_D+\Omega_{TO}}(t)]\times [{\bf r_i+u_i}] \right )^2.
$$
The change in energy is thus
$$
\Delta E\approx NL_d\hbar^2/(2m_{kink}L_d^2)
+m\sum_i(\Omega_D+\Omega_{TO})^2 [0.5\delta u_{i}(t)^2
+\delta u_{i}(t)(r_i+u_{i})].
$$
We now look at $\Delta F$, the change in F as a kink wave is created.

In general, $r>>u(r),$ after discarding contributions
with zero time averages, we obtain
\beq
\Delta F\approx NL_d\hbar^2/(2m_{kink}L_d^2)
-m\sum_i[\Omega_D^2 \delta u_{li}(t)r_i
+2\Omega_{TO}(t) \Omega_D r_i\delta u_{li}(t)
+ \Omega_{TO}(t) r_i\partial u_{ti}/\partial t] \label{dfd}
\eeq

The last term is the same as in the AC case.
Since $|\Omega_D|>>|\Omega_A|,$ the term
$2\Omega_{TO}(t) \Omega_D r_i\delta u_{li}(t)$ is much smaller than
$\Omega_D^2 \delta u_{li}(t)r_i $ and will be ignored.
In this sum there is now a new driving term 
$
-m\sum_i  \Omega_D^2 \delta u_{li}(t)r_i
$ 
that couples to a constant change of position of the kinks.
Consider, for example, the wave function  
$\phi(z)\propto [1+\sin (2\pi z/L_d)]$
which is a linear combination of the state $|k=0>$ 
and the states $|k=\pm 2\pi/L_d>$. This state has a constant 
shift in the kink position. 
Once this state is created, the oscillating Hamiltonian can 
couple the states $\phi$ to an oscillating state such as 
$\phi'(z)\propto \cos\omega_{TO} [1+\sin (2\pi z/L_d)].$

The displacement
$\delta u_{li}$ is of the order $L_d,$ the new term is of the 
order of magnitude
$
-m \Omega_D^2L_d L_m^2/a^2. 
$ 
Substituting this into eq. (\ref{dfd}) and setting $\Delta F=0$
we thus arrive at a critical DC angular velocity of 
the order of magnitude
\beq
\Omega_D^{crit}\approx [\Omega_A^{crit}v_0/L_d]^{1/2}
\eeq
From this we obtain an estimate of $\Omega_D^{crit}$ of the order
of rad/s, the same order of magnitude as the expeimental results.

(ii) We have considered an alternative mechanism due to the onset
of the tunnelling
of a kink wave across the node in the dislocation network. 
We find a critical angular velocity given by
\beq
\Omega_D^{crit}=[\Omega_A^{crit} \omega_{TO}2\pi]^{1/2}.\label{jt}
\eeq
This critical velocity is
of the order of 10 rad/s, a little higher than the experimental value.
For this mechanism, $\Omega_D^{crit}$ is a function of the torsional
oscillator frequency whereas this is not true with the other mechanism.
We explain this next.

We have investigated this by modelling our calculation 
along the lines 
similar to the Josephson effect with the node of the
network modelled as the insulating barrier .
Under the oscillating rotation, due to the centrifugal force
there is an effective "potential" 
$q\Delta V\approx m L_m^2 L_d\Omega^2/a$ driving the kinks of the
dislocations across the node. 
$\Omega$ and hence $q\Delta V$ contains both a
DC contribution
$q\Delta V_D\approx m L_m^2 L_d\Omega_D^2/a$ 
and an AC part
$q\Delta V_A\approx m L_m^2 L_d2\Omega_D\Omega_A\cos(\omega_At)/a.$ 
As we learned from the Josephson equations\cite{Feynman}, a 
current of kinks can develop across the node that
contains a term given by $$J=q\Delta V_A\sin\omega_{TO}t\cos(
\delta_0+q\Delta V_Dt/\hbar)/(\hbar\omega_{TO}),$$
where$\delta_0$ is a constant phase difference.
The critical velocity is reached when a DC component 
of the current is developed
across the junction. This happens when the quantum energy
associated with the oscillation frequency 
$\hbar\omega_{TO}$ is equal to the effective potential applied due to 
the centrifugal force $q\Delta V_D$. 
We obtain a critical DC angular frequency
given by eq. (\ref{jt}).
We close this paper with other issues that we have considered.

As is mentioned above,
in general, the axis of rotation is not parallel to the axis of the
dislocation.
The crystal orientation can be specified
by two Euler angles $(\theta, \Phi)$
with respect to the rotation axis. (The third
angle corresponds to the angle of rotation).
The actual displacement
from the dislocation motion which contributes to the
kinetic energy of the particles
should be ${\bf R u(R^{-1}r})$ where {\bf R} is the rotaion
matrix that can be specified by the Euler angles. 
We have explicitly computed this quantity and
verified that our results are as
expected. More precisely we find that
$
\int d^2r {\bf r\times R\partial_{x'}u(R^{-1}r)}=
 0.5  \cos 2\Phi  F(\theta)$,
$F(\theta)=\int d^2r[
 (1-2s) (x^4 \cos \theta - y^4\cos^3 \theta ) 
+  (\cos^{2} \theta -1) ( 3 - 2s)\cos \theta  y^2{x}^{2} ] 
/[(s-1)( \cos^2 \theta  y^2+x ^2 ) ^2]
$
Similarly
 we obtain
$
\int dr {\bf r\cdot R\partial_{x'}u(R^{-1}r)}=\sin(2\Phi)G(\theta)
$
where
$
G=-[ {x}^{2}  \cos^{2}  ( \theta
 )  
- {y}^{2}]/[{y}^{2}+ \cos^{2} ( \theta )   {x}^{2}]
$


We were also concerned about possible changes in the phonon
dispersion due to the rotation
and its effect on the energetics of the system.
We find that the dominant contribution to the energy
change is given by
$E_{2sd}
=0.25 (\hbar/N)\sum_{k,j}|{\bf
\Omega\times e_j}|^2(2n_{kj}+1)/\omega_k.$
where $k$, $j$ specifies the wave vector and branch index
of the phonons with frequency $\omega_k$, polarization ${\bf e}_j$
and occupation number $n_{kj}.$ Since the 
phonon frequencies are of the order of $10^{12}/sec$ and 
$\Omega$ is less than $rad/s$, these corrections are small.

In summary
we apply our recently developed model of a Bose condensate
of quantum kink wave in solid He4
to understand recent experimental results of the
citical velocities and the effect
of the steady and oscillating rotations. 
Estimates
of the critical velocities and the change in the Q value of 
the trosional oscillator 
with no adjustable parameters are 
of the same order of magnitude as the experimental results.
Their functional dependence on system parameters is discussed.
We thank Norbert Mulders for helpful discussions.


\begin{thebibliography}{99}
\bibitem{chan}
E. Kim and M. W. H. Chan, Science 305 1941 (2004).
\bibitem{x}
C. A. Burns, N. Mulders, L. Lurio, M. H. W. Chan, A. Said, C. N. Kodituwakku
and P. M. Platzman, Phys. Rev. B78, 224305 (2008).
\bibitem{jb}
J. Day and J. Beamish, Nature (London) 450, 853 (2007).
Yu. Mukharsky, A. Penzev, E. Varoquaux, Phys. Rev. B 80, 140504, (2009).
X. Rojas, C. Pantalei, H. J. Maris, S. Balibar, JLTP 158, 478, (2010);
J. Day, O. Syshchenko and J. Beamish, Phys. Rev. Lett. 104, 075302 (2010).
O. Syshchenko, J. Day and J. Beamish, Phys. Rev. Lett. 104, 195301 (2010).
\bibitem{balibar}
S. Sasaki, R. Ishiguro, F. Caupin, H. J. Maris, and S. Balibar,
Science 313, 1098 (2006).
\bibitem{reppy}
A. S. C. Rittner and J. Reppy, Phys. Rev. Lett. 98, 175302 (2007).
\bibitem{toner}
J. Toner, Phys. Rev. Lett. 100, 035302 (2008).
\bibitem{lagb}
L. Pollet, M. Boninsegni, A. B. Kuklov, N. V. Prokofev, B. V. Svistunov and M. Troyer, Phys. Rev. Lett. 98, 135301 (2007).

\bibitem{Choi}
H. Choi, T. Takahashi, E. Kono, E. Kim, Science 330, 1512 (2010).
\bibitem{Kubota}
M. Yagi, A. Kitamura, N. Shimizu, Y. Yasuta, M.  Kubota, 
J Low Temp Phys 162: 492; 162, 754 (2011).
\bibitem{kwave}
S. T. Chui, Phys. Rev. B82, 014519 (2010).
\bibitem{LL} L. D. Landau and E. M. Liftshitz, 
"Statistical Physics" 2nd Ed., p.99
Addison-Wesley, Reading, MA. (1969).
\bibitem{kojima}
Y. Aoki, J. C. Graves, and H. Kojima, PRL 99, 015301 (2007);
Y. Aoki, M. C. Keiderling, and H. Kojima,
PRL 100, 215303 (2008).

\bibitem{ra} 
There is an additional term
$\int d{\bf r}\rho [{\bf \Omega_{TO}(t) \times \delta u}]^2$
This term is of the order of $ln(L_m/a)(\Omega_{TO}
\times \delta c)^2$
Since $\Omega_{TO}\approx 10^{-3}/second,$
the ratio  between the term kept and this term
is $10^{-6} ln(L_m/a)[\delta c/second /v_0]^2.$ 
We take $v_0\approx 10^{7} nm/second $ and $\delta c < 1 nm.$ This ratio
is of the order of $10^{-20}$
and is much smaller than the first term 
It will be neglected from now on.

%
%
%
%

%
%
\bibitem{uxy} 
$u_x=
\arctan (y'/x') + 0.5 x' y'/[(  x'^{2}+ y'^{2} ) 
( 1-\sigma ) ]
$, 
$
u_y= - 0.5\, [  0.5\, ( 1-2\,\sigma ) \ln  (  x'^{2}
+ y'^{2} ) + x'^{2}/( x'^{2}+ 
y'^{2})] / ( 1-\sigma ) 
$
where $x'=x-c_x,$ $y'=y-c_y,$ $\sigma$ is the Poisson ratio.
\bibitem{eqn}
See eq. (B1) of ref. (\cite{kwave}) 
\bibitem{gold}
H. Goldstein, 
"Classical Mechanics", section 4.4 
\bibitem{Feynman} See, for example, vol.3 Lectures on Physics. R. P. Feynman, p. 21-16.

\end{thebibliography}
\end{document}